\documentclass[pra,twocolumn,a4paper,showpacs,floatfix]{revtex4}

\usepackage{graphicx}
\usepackage{amssymb}

\begin{document}

\title{The three-dimensional Anderson model of localization with binary
  random potential}

\author{I~V~ Plyushchay\dag\footnote[3]{innapl@univ.kiev.ua},
  R~A~R\"{o}mer\ddag\footnote[4]{r.roemer@warwick.ac.uk} and
  M~Schreiber\P\footnote[6]{schreiber@physik.tu-chemnitz.de}}

\affiliation{\dag Department of Physics, Kiev Taras Shevchenko
National University, Volodymyrska~st.~64, Kiev, 01033 Ukraine}

\affiliation{\ddag Department of Physics, University of Warwick,
Coventry, CV4 7AL, United Kingdom}

\affiliation{\P Institut f\"{u}r Physik, Technische
Universit\"{a}t, D-09107 Chemnitz, Germany}

\date{$Revision: 1.37 $, compiled \today}

\begin{abstract}
  We study the three-dimensional two-band Anderson model of localization
  and compare our results to experimental results for amorphous
  metallic alloys (AMA). Using the transfer-matrix method, we identify
  and characterize the metal-insulator transitions as functions of Fermi
  level position, band broadening due to disorder and concentration of
  alloy composition. The appropriate phase diagrams of regions of
  extended and localized electronic states are studied and qualitative
  agreement with AMA such as Ti-Ni and Ti-Cu metallic glasses is found.
  We estimate the critical exponents $\nu_W$, $\nu_E$ and
  $\nu_x$ when either disorder $W$, energy $E$ or concentration $x$ is
  varied, respectively. All our results are compatible with the
  universal value $\nu\approx1.6$ obtained in the single-band Anderson
  model.
\end{abstract}


\maketitle

\section{Introduction}
\label{sec-intro}

Amorphous metallic alloys (AMA) offer the possibility of
continuously changing their composition while at the same time
avoiding structural phase transformations. Thus they allow for
systematic studies of their physical properties within a single
phase as temperature and other external control parameters, e.g.,
pressure, are varied. Many such investigations (see, e.g.,
\cite{Has83}) have been devoted to the investigation of the
electrophysical properties (in particular, electrical resistivity)
of AMA; these studies have revealed considerable differences in
the behavior of the electrical resistivity $\varrho$ of AMA in
comparison with that of their crystalline counterparts.

The transport properties of transition-metal-based AMA are of
special interest because of unique physical anomalies, unexplained
by conventional transport theory, that exist in these materials.
Among them are the negative temperature coefficient of
resistivity, the Mooij correlation of resistivity and
thermoelectric power \cite{Moo73} --- the higher the resistivity
the lower is the temperature coefficient of the resistivity ---,
the resistivity saturation, the sign reversal of the Hall
coefficient, a negative magneto-resistance, and the breakdown of
Mattiessen's rule. For a review, we refer the reader to ref.\
\cite{LeeR85}.
The first attempt for a theoretical understanding of the
electronic transport in noncrystalline materials was the Ziman
theory and its extensions \cite{Has83}. However, the applicability
of the Ziman diffraction model for strongly disordered and thus
high-resistivity amorphous alloys is questionable
\cite{CotM82,MeiC84}.  In addition, the Ziman theory cannot
explain, e.g., that the changes of the resistivity $(\varrho_{\rm
liquid}-\varrho_{\rm solid})/\varrho_{\rm solid}$ at the melting
point are only $0.01$ and $0.09$ for Fe and Co, respectively
\cite{Has83}. Thus it appears that the resistivity of these two
elements depends only weakly on their atomic structure factor
$S(k)$.

At low temperature $T$, an even more significant difference
between the behavior of crystals on the one hand and disordered
solids on the other hand is seen: sufficiently strong disorder can
give rise to a transition of the transport properties from
conducting behavior with resistance $R > 0$ to insulating behavior
with $R=\infty$ as $T\rightarrow 0$ as was pointed out by Anderson
in 1958 \cite{And58}, This phenomena is called the disorder-driven
metal-insulator transition (MIT) \cite{BelK94,KraM93,LeeR85} and
it is characteristic to non-crystalline solids. The mechanism
underlying this MIT was attributed by Anderson not to a finite gap
in the energy spectrum which is responsible for an MIT in band gap
or Mott insulators \cite{AshM76}.  Rather, he argued that the
disorder will lead to interference of the electronic wave function
$\psi({\bf r})$ with itself such that it is no longer extended
over the whole solid but is instead confined to a small part of
the solid. This {\em localization} effect excludes the possibility
of diffusion at $T=0$ so that the system is an insulator.
A highly successful theoretical approach to this disorder-induced
MIT was put forward in 1979 by Abrahams {\em et al.}
\cite{AbrALR79}. This ``scaling hypothesis of localization''
details the existence of an MIT for non-interacting electrons in
three-dimensional disordered systems at zero magnetic field and in
the absence of spin-orbit coupling.

These ideas have also been applied to the analysis of AMA
transport properties \cite{Imr79}. They provide a background for
an explanation of the Mooij correlation \cite{ParSK01} and were
successful in the description of the transport properties of
amorphous semiconductors \cite{Mad78}.
Many papers have since  discussed the influence of localization
(or quantum interference effects) on the transport properties of
AMA \cite{HicGH86, BogH90, HowHM88, BieFGO84}. It was shown
\cite{SeeLHM99} that Anderson localization is responsible for
regions of high resistivity.  The importance of quantum
corrections in analyzing the conductivity of so-called highly
resistive ($\varrho>150\mu\Omega$cm) metallic glasses was
demonstrated \cite{HowG86}. There are two sources of these
corrections: the disorder-induced ``localization effect'' and the
electronic ``interaction effect''. The former caused the observed
higher temperature resistivity. Furthermore, it has been argued
\cite{RosHPK97, Tsu86} that for amorphous metals localization
effects are valid even at room temperature.

In a recent study \cite{NakPSZ00} the influence of disorder on the
transport properties of Ti-Ni and Ti-Cu metallic glasses was analyzed
within the Anderson model of localization. The results
elucidated qualitatively the correlation between the observed details of
the electronic structure and the behavior of resistivity vs.\
temperature for this binary AMA.  It was shown that the temperature
coefficient of resistivity depends on the position of the Fermi level
$E_{\rm F}$ and on the (assumed) position of the mobility edge $E_{\rm
  c}$. The latter separates extended (conducting) states from
localized (insulating) states \cite{KraM93,GruS95}.  Unfortunately, the
value of $E_{\rm c}$ had to be inferred from the hypothesis that
the two subbands of the valence band of the binary alloy have
tails with localized electronic states like the tails of the usual
one-band Anderson model \cite{And58}. Evidently, a calculation of
the position of the mobility edge $E_{\rm c}$ should be carried
out for the case of two subbands. This is what we intend to do in
the present manuscript.  In particular, we will investigate
whether localized electronic states exist in the central region
between the two subbands as assumed in \cite{NakPSZ00}.
Furthermore, we show how the concentration parameter, absent in
the usual single-band Anderson model, influences the position of
the mobility edges.

\section{Numerical approach}

\subsection{The binary Anderson model of localization}
\label{sec-BAM}

According to the photo-emission data and theoretical estimates
\cite{Haf91,NakPSZ00}, the valence band of a binary,
transition-metal-based amorphous alloy can be assumed to be a
superposition of two valence bands $A$ and $B$. In order to model
this within the Anderson model, we use the standard Anderson
Hamiltonian \cite{And58}
\begin{equation}
  \label{eq-Hand}
  {\bf H} = \sum_{i} \varepsilon_{i} | i \rangle\langle i | + \sum_{i \ne
    j} t_{ij} | i \rangle\langle j | \quad
\end{equation}
with orthonormal states $| i \rangle$ corresponding to electrons located
at sites $i=(x,y,z)$ of a regular cubic lattice with periodic boundary
conditions. The hopping integrals $t_{ij}$ are non-zero only for nearest
neighbors and we set the energy scale by choosing $t_{ij}=t=1$.
The two-subband structure observed in the experiments is
incorporated into the potential energies $\varepsilon_{i}$.  The
randomness is modeled by
\begin{enumerate}
\item independent random variations $\varepsilon_{i(A)} \in
  [\varepsilon_{A}-W_A/2, \varepsilon_{A}+W_A/2]$ and $\varepsilon_{i(B)} \in
  [\varepsilon_{B}-W_B/2,\varepsilon_{B}+W_B/2]$,
\item random spatial distribution of potential energies
$\varepsilon_A$ and $\varepsilon_B$ within the cubic lattice.
\end{enumerate}
The mean values $\varepsilon_{A}$ and $\varepsilon_{B}$ are chosen
according to the central energies of the two subbands. The
parameters $W_A$ and $W_B$ specify the disorder strength in each
energy band and the random variation of $A$ and $B$ sites models
the compositional disorder of the binary alloy. Note that with
this choice of parameters, the band edges of the subband $A$ in
the limit of large system size are given by $\varepsilon_{A} - 6
-W_A / 2$, $\varepsilon_{A} + 6 + W_A / 2$ and similarly for the
other subband with $A\rightarrow B$.

In summary, the model is described by the following parameters
\begin{enumerate}
\item $\varepsilon_{AB}= \varepsilon_{B} - \varepsilon_{A}$, the
distance between the centers of the two bands (we choose
$\varepsilon_{B}>\varepsilon_{A}$ so that the $A$ band is the
lower one), \item $x_{A}$, the concentration of $A$ sites (of
course, then $x_{B}= 1 - x_{A}$ is the concentration of the $B$
sites), \item $W_A$ and $W_B$, the widths of the box distributions
of the potential energy.
\end{enumerate}

\subsection{Transfer-matrix method}

Since we are interested in the position of the mobility edges and thus
the localization lengths, we use the transfer-matrix method (TMM)
\cite{KraM93,PicS81a,MacK81} for the investigation of model
(\ref{eq-Hand}).  The localization length $\lambda$ describes the
exponential decay of the wave function and we compute it using TMM for
quasi-1D bars of cross section $M\times M$ and length $L \gg M$. As is
customary in the TMM approach, the stationary Schr\"odinger equation
${\bf H}\Psi=E\Psi$ is rewritten in the recursive form
\begin{equation}\label{eq-recursion}
  {\Psi_{i+1} \choose \Psi_{i}} =
  {\left(
      \begin{array}{cc}
        E{\bf 1} -{\bf H}_i & {\bf -1} \\
        {\bf 1} & {\bf 0}
      \end{array}
    \right)}
  {\Psi_{i} \choose \Psi_{i-1}} =
  {\bf T}_i {\Psi_{i} \choose \Psi_{i-1}} .
\end{equation}
$\Psi_{i}$, ${\bf H}_i$, and ${\bf T}_i$ are wave function,
Hamiltonian matrix, and transfer matrix of the $i$th slice of the
bar, respectively. Unit and zero matrices are denoted by ${\bf 1}$
and ${\bf 0}$. Given an initial condition ${\Psi_{1} \choose
\Psi_{0} }$ equation (\ref{eq-recursion}) allows a recursive
construction of the wave function in the bar geometry by adding
more and more slices.  $\lambda(M,w)$ is then obtained from the
smallest Lyapunov exponent of the product ${\bf T}_L {\bf T}_{L-1}
\cdots {\bf T}_2 {\bf
  T}_1$ of transfer matrices \cite{MacK83}, where the length $L$ of the
bar is increased until the desired accuracy of $\lambda$ is
achieved. With increasing cross section of the bar the reduced
localization length $\Lambda_M(w)=\lambda(M,w)/M$ decreases for
localized states and increases for extended states.  Thus it is
possible to determine the critical parameter $w_{\rm c}$ at which
$\Lambda_M$ is constant as a function of the varied parameter $w$
--- e.g., $E$, $\varepsilon_{AB}$, $x_{A}$, $W_A$, or $W_B$ ---
from plots of $\Lambda_M$ versus $M$.

\subsection{Finite-size scaling}
\label{sec-FSS}

The MIT in the Anderson model of localization is expected to be a
second-order phase transition \cite{BelK94,AbrALR79}. It is
characterized by a divergent correlation length
$\xi_\infty(w)=C|w-w_{\rm c}|^{-\nu}$, where $\nu$ is the critical
exponent, $C$ is a constant \cite{KraM93}, $w$ is any of the
external control parameters given above and $w_{\rm c}$ is its
critical value at which the MIT occurs. To construct the
correlation length of the {\em
  infinite} system $\xi_\infty$ from finite-size data $\Lambda_M$
\cite{ZamLES96a,KraM93,PicS81a,MacK81}, the one-parameter scaling
hypothesis \cite{Tho74} is employed,
\begin{equation}
  \label{eq-ScalFunc}
  \Lambda_M=f(M/\xi_\infty) \quad .
\end{equation}
All values of $\Lambda_{M}(w)$ are expected to collapse onto a
single scaling curve $f$, when the system size is scaled by
$\xi_{\infty}(w)$. In a system with MIT such a scaling curve
consists of two branches corresponding to the localized and the
extended phase. One might determine $\nu$ from fitting
$\xi_\infty$  by a finite-size scaling (FSS) procedure
\cite{MacK83}. But a higher accuracy can be achieved by fitting
directly the raw data \cite{MacK83}. We use fit functions
\cite{SleO99a} which include two kinds of corrections to scaling:
(i) nonlinearities of the $w$ dependence of the scaling variable
and (ii) an irrelevant scaling variable with exponent $-y$.
Specifically, we fit
\begin{equation}
  \label{eq-SlevenRenorm2}
  \Lambda_M=\tilde{f}_0(\chi_{\rm r} M^{1/\nu})+M^{-y}
  \tilde{f}_1(\chi_{\rm r} M^{1/\nu}) \quad ,
\end{equation}
\begin{equation}
  \label{eq-SlevenRenorm3}
  \tilde{f}_n=\sum_{i=0}^{n_{\rm r}} a_{ni} \chi_{\rm r}^i M^{i/\nu}
  \quad, \quad \chi_{\rm r}(\omega)=\omega+\sum_{n=2}^{m_{\rm r}} b_n \omega^n
\end{equation}
with $\omega=|w_{\rm c}-w|/w_{\rm c}$ and expansion coefficients
$a_{ni}$ and $b_n$. Choosing the orders $n_{\rm r}$ and $m_{\rm r}$ of
the expansions larger than one, terms with higher order than linear in
the $w$ dependence appear.  This allows to fit a wider $w$ range around
$w_{\rm c}$ than with the previously used linear fitting \cite{Mac94}.
The linear region is usually very small. The second term in equation
(\ref{eq-SlevenRenorm2}) describes the systematic shift of the crossing
point of the $\Lambda_M(w)$ curves \cite{Mac94,SleO99a}.

In the present case of the two-band Anderson model, we can in principle
have an MIT as a function of
\begin{enumerate}
\item energy $E$ for fixed $W_A$, $W_B$, $\varepsilon_{AB}$ and $x_{A}$,
\item disorder strengths $W_{A}$, $W_B$ for fixed $E$,
  $\varepsilon_{AB}$ and $x_{A}$,
\item concentration $x_{A}$ for fixed $E$, $\varepsilon_{AB}$ and $W_A$,
  $W_B$.
\end{enumerate}
Due to universality, we expect the corresponding critical exponents
$\nu_E$, $\nu_{W_A}$, $\nu_{W_B}$ and $\nu_x$ to be the same.
Additionally for each control parameter we test whether the fitted
values of $w_{\rm c}$ and $\nu$ are compatible when using different
expansions of the fit function, {i.e.}, different orders $n_{\rm r}$ and
$m_{\rm r}$ \cite{MilRS00}.

\section{Results and discussion}

For the determination of the mobility edge \cite{GruS95}, we have
determined the localization lengths from TMM with an error equal to or
less than $1\%$ for cross-sections up to $M=12$.  For the computation of
the critical exponents, we have used the available $1\%$ data and
generated additional data with $0.05\%$ error up to at most $0.1\%$ at
selected points in the phase diagrams for high-precision estimates.

\begin{figure}
\begin{center}
  \includegraphics[width=0.95\columnwidth]{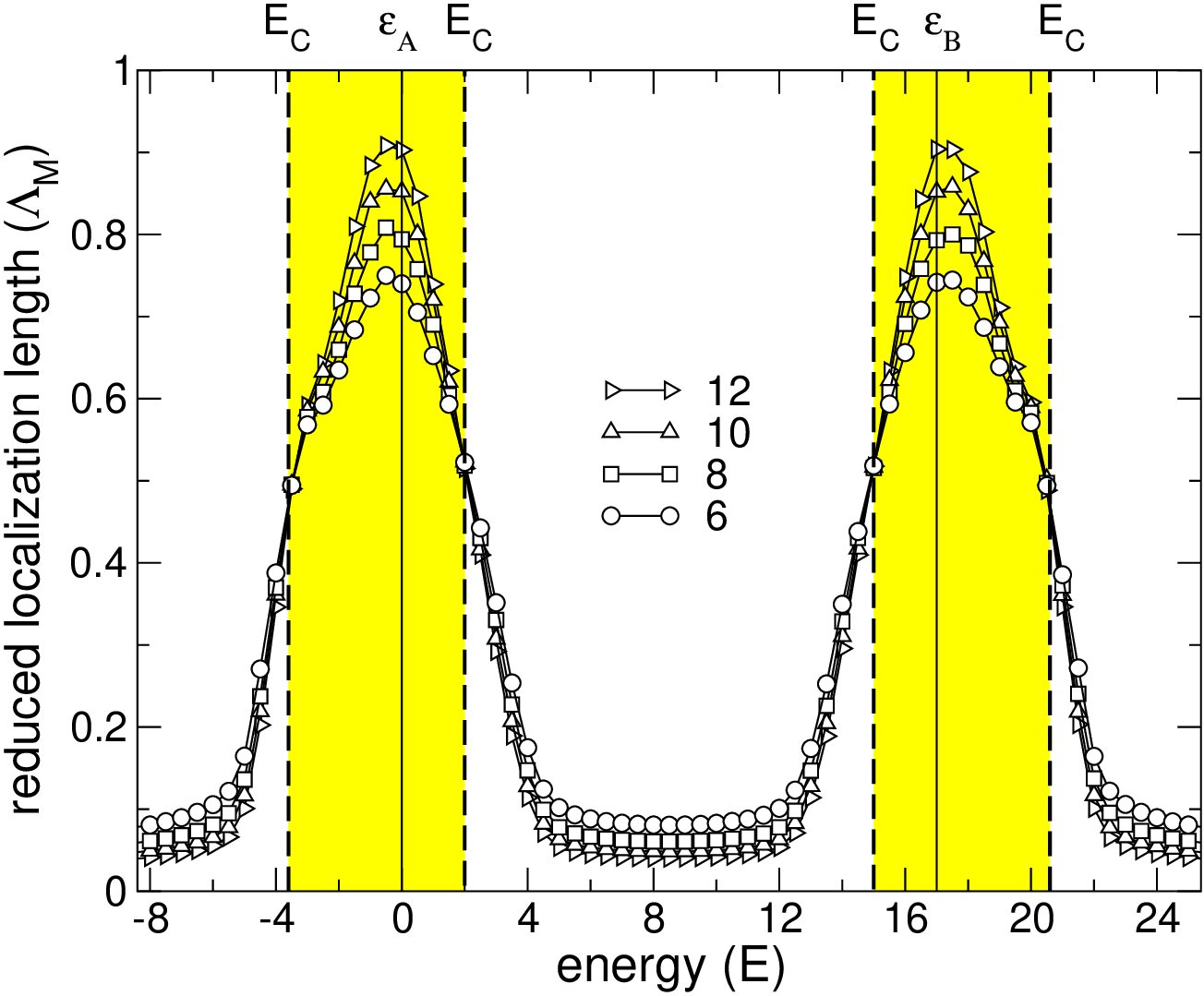}
\caption{\label{fig-Lambda_E}
  Dependence of the reduced localization length $\Lambda_M$ on energy
  $E$ for different system sizes $M=6, 8, 10$ and $12$ at
  $\varepsilon_{AB}=17$, $x_{A}=0.5$ and $W_A=W_B=5$. The two shaded
  regions indicate extended states, the dashed lines the
  position of the 4 mobility edges, and the vertical solid lines the band
  centres of the subbands. No error bars are shown because the accuracy
  of the $\Lambda_M$ values is better than the size of the symbols.}
\end{center}
\end{figure}

In Fig.~\ref{fig-Lambda_E}, we show a typical dependence of the
reduced localization length $\Lambda_M$ on energy for different
system sizes $M$. The two peaks correspond roughly to the
positions of the two subbands at $\varepsilon_{A}$ and
$\varepsilon_{B}$, respectively. This is in accordance with the
theoretical and experimental results \cite{NakPSZ00,Haf91}
mentioned in the introduction as motivation of our studies. Note
that for the chosen parameters, the density-of-states limit given
in section \ref{sec-BAM} for the highest energy state in the lower
subband coincides with the smallest possible energy state in the
upper subband at $\varepsilon_{A} + 6 + W_A/2 = \varepsilon_{B} -
6 - W_B/2$.
At the positions indicated by the  dashed lines in
Fig.~\ref{fig-Lambda_E}, we see a reversal in the systematic size
dependence of the reduced localization lengths. Therefore these
indicate the mobility edges and we observe localized electronic
states at the band edges   and in the central region between the
$A$ and $B$ subbands. We remark that the $\Lambda_M(E)$ curves
demonstrate a peculiar feature: the positions of the maximum
values of $\Lambda_M$ do not coincide with the central energies
$\varepsilon_{A}$ and $\varepsilon_{B}$  of the two subbands. Thus
although the density of states between the two bands is
appreciable, we observe the surprising fact that the region of
extended states in each band has shrunk.

\begin{figure}
\begin{center}
  \includegraphics[width=0.95\columnwidth]{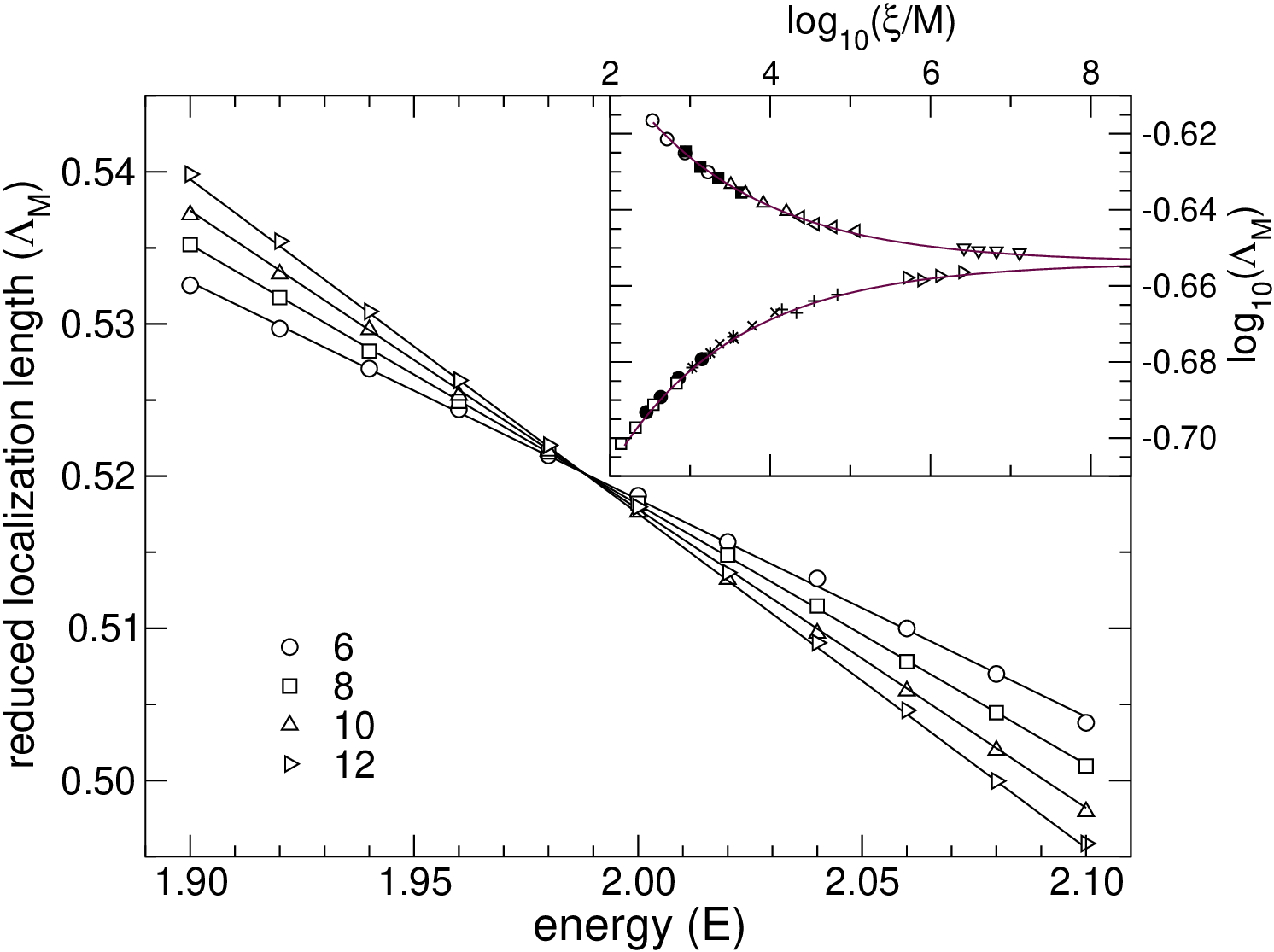}
\caption{\label{fig-fss-Lambda_E}
  FSS plot of the reduced localization length $\Lambda_M$ near the
  mobility edge $E_{\rm c}\approx 2$ of Fig.\ \protect
  \ref{fig-Lambda_E}. The solid lines are fits of the data according to
  equations (\ref{eq-SlevenRenorm2}) and (\ref{eq-SlevenRenorm3}) with
  $n_{\rm r}=1$ and $m_{\rm r}=1$.  The inset shows the scaling
  function corresponding to the fit and the $\Lambda_M$ data with
  symbols $\circ, \blacksquare, \triangle, \triangleleft, \triangledown,
  \triangleright, +, *, \bullet, \square$ denoting energies $1.9, 1.92,
  1.94, \ldots, 2.08, 2.10$, respectively.}
\end{center}
\end{figure}

In Fig.~\ref{fig-fss-Lambda_E} we show high precision data at the
upper mobility edge of the $A$ subband at $E_{\rm c}\approx 2$.
Again the accuracy of the data is so high that error bars would be
smaller than the symbols.
Data and FSS analysis of similar quality will be used when estimating
critical exponents in the following.

\subsection{Energy-disorder phase diagram}
\label{sec-en-dis}

\begin{figure*}
\begin{center}
  \includegraphics[width=0.95\columnwidth]{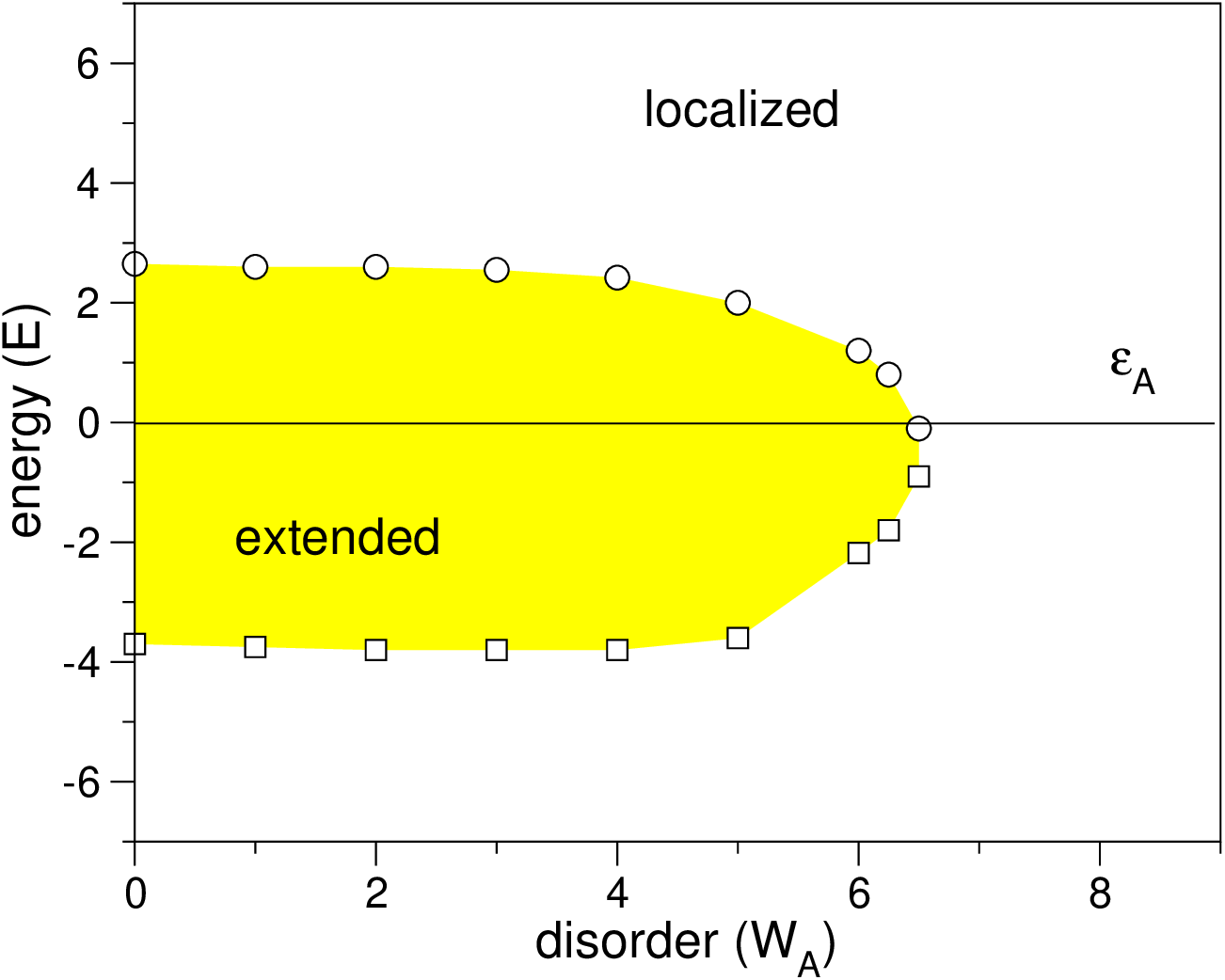}
\caption{\label{fig-phase-E_WA}
  The energy-disorder phase diagram for $\varepsilon_{AB}=17$ and
  $x_{A}=0.5$. The shaded region indicates extended states and the
  horizontal line denotes the value of $\varepsilon_A$. Here and in the
  following phase diagrams, $\square$ and $\circ$ indicate lower and
  upper mobility edges for the $A$ band, and, if included, upper and
  lower mobility edges for the $B$ band.}
\end{center}
\end{figure*}

Let us now investigate how the positions of the mobility edges
change when  $W_A$ and $W_B$ are changed.  We set
$\varepsilon_{AB}=17$ and $x_{A}=x_{B}=0.5$.  For convenience we
choose $W_A\equiv W_B$. This leads to a symmetry for the energy
dependences of $\Lambda_M(E)$ between lower and upper subband as
shown in Fig.~\ref{fig-Lambda_E}. Consequently, the phase diagram
is symmetric with respect to $\varepsilon_{A} +
\varepsilon_{AB}/2$. We find extended electronic states in the
vicinity of $\varepsilon_{A}$ and for  $W_A \leq W_{\rm c}\approx
6.5$ (instead of $16.5$ as for the usual Anderson model) as
presented in Fig.~\ref{fig-phase-E_WA}.
For larger $W_A$ or larger $|E-\varepsilon_{A}|$, the states are
localized.

Next, we keep $W_A=4$ fixed and vary $W_B$ from 0 to 8. We find
for $\varepsilon_{AB}=17$ that the position of mobility edges in
the subband $A$ is only slightly modified. The same behavior is
observed for $\varepsilon_{AB}=8$, when the overlap of the two
subbands is large.

\subsection{Energy-$\varepsilon_{AB}$ phase diagram}

\begin{figure}
\begin{center}
  \includegraphics[width=0.95\columnwidth]{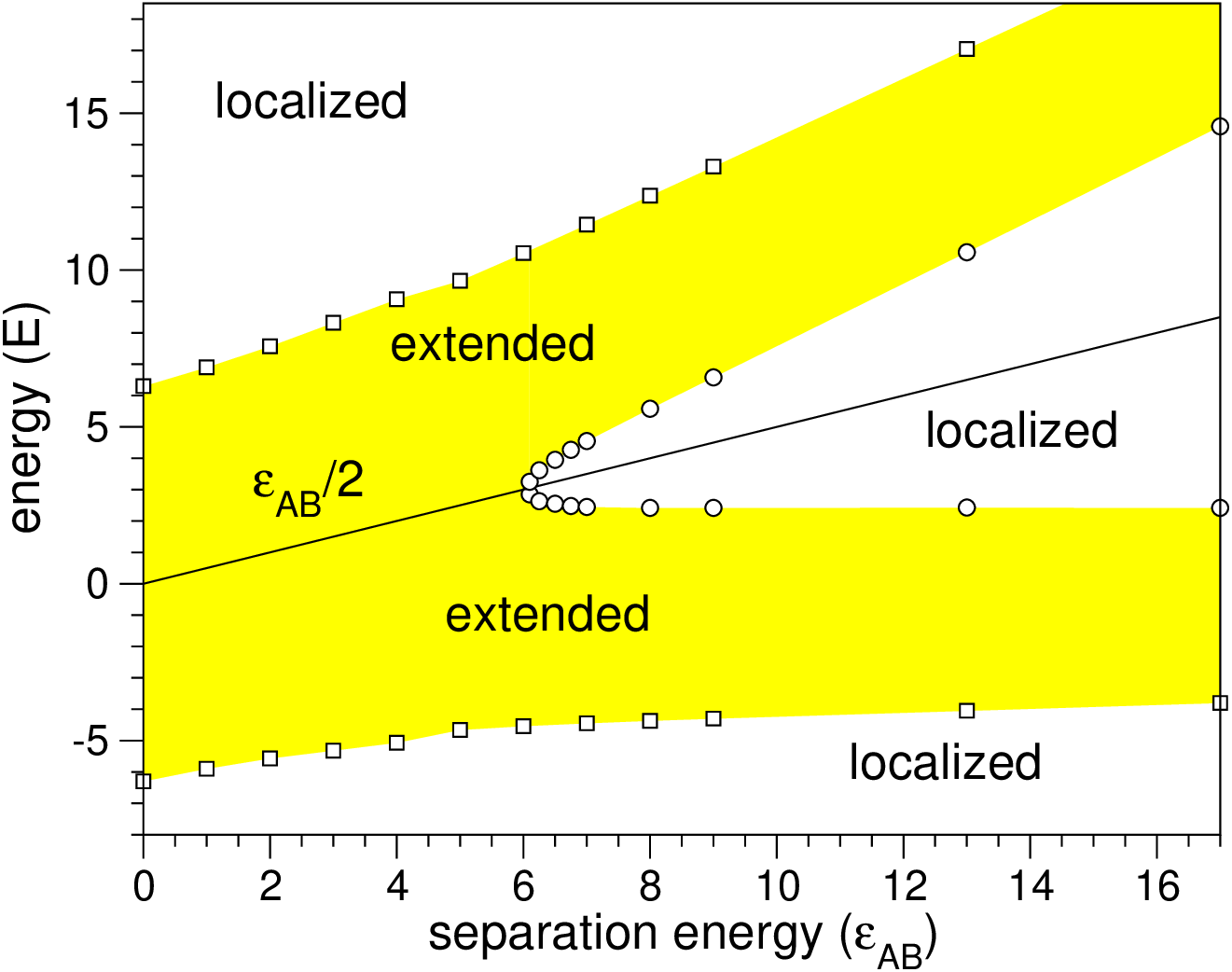}
\caption{\label{fig-phase-E_EAB}
  The energy-$\varepsilon_{AB}$ phase diagram for $W_A=W_B=4$ and
  $x_{A}=0.5$. The shaded region indicates extended states. The upper
  ($B$) and lower ($A$) bands are symmetric with respect to
  $\varepsilon_{AB}/2$ and the $B$ band has been constructed using this
  symmetry.}
\end{center}
\end{figure}

Keeping $x_{A}=0.5$ and fixed disorder $W_A=W_B=4$, we now vary
the subband spacing $\varepsilon_{AB}$. From an experimental point
of view, this is the difference between the atomic constituents of
the alloy
 or, more precisely, between their ionization energies.
For a variety of $\varepsilon_{AB}$ values we have obtained
$\Lambda_M$ curves which are qualitatively very similar to
Fig.~\ref{fig-Lambda_E}. However, for $\varepsilon_{AB}\leq 6$, no
localized electronic states in the central region can be found. In
Fig.~\ref{fig-phase-E_EAB}, we show the  mobility edges for
different $\varepsilon_{AB}$.
We note that the mobility edge for the usual Anderson model equals
approximately $\pm 6.2$ at disorder $W=4$ \cite{KraM93,BulSK87}.
This agrees with the result for $\varepsilon_{AB}=0$ in Fig.
~\ref{fig-phase-E_EAB}.

\subsection{Energy-concentration phase diagram}
\label{sec-en-conc}

\begin{figure}
\begin{center}
  \includegraphics[width=0.95\columnwidth]{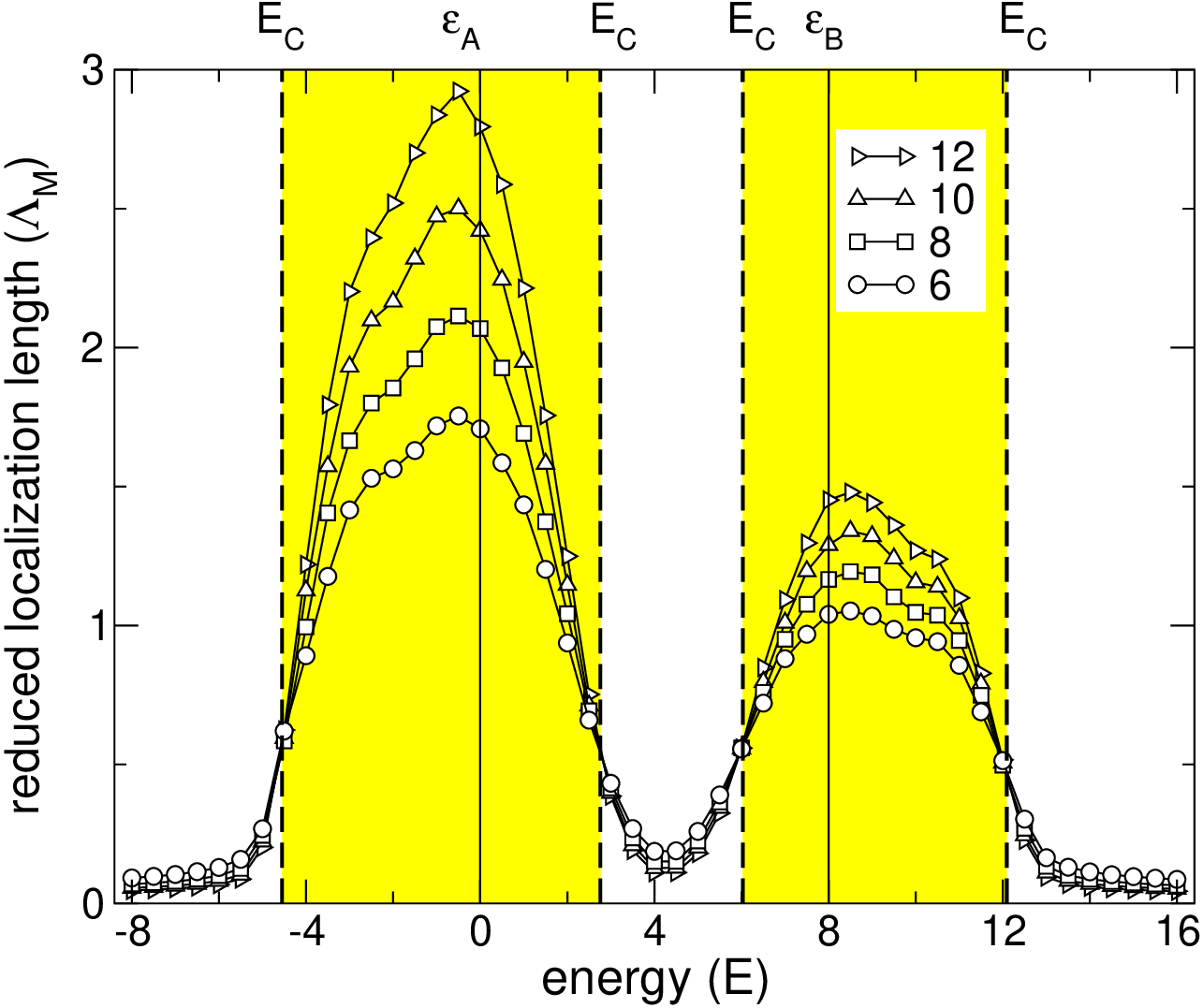}
\caption{\label{fig-Ab-Lambda_E}
  The energy dependence of $\Lambda_M$ for a majority of $A$ sites
  (concentration $x_{A}=0.55$). System sizes are $M=6, 8, 10$ and $12$,
  $\varepsilon_{AB}=8$, and $W_A=W_B=4$. The two shaded regions
  indicate extended states. The solid and dashed vertical lines denote
  the band energies $\varepsilon_{A}$ and $\varepsilon_{B}$ and the
  position of the $4$ mobility edges, $E_{\rm c}$, respectively.}
\end{center}
\end{figure}

\begin{figure}
\begin{center}
  \includegraphics[width=0.95\columnwidth]{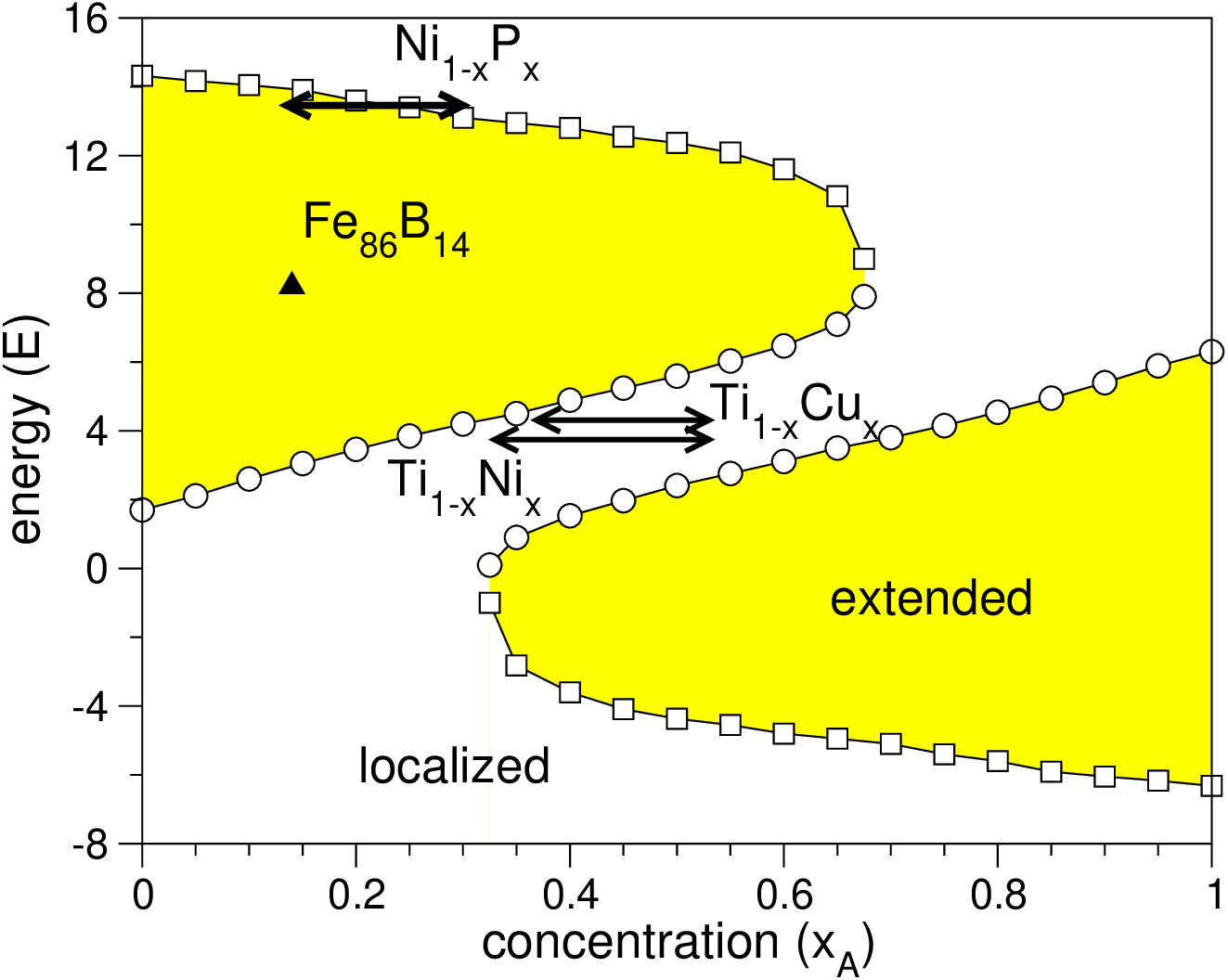}
\caption{\label{fig-phase-E_CA}
  The energy-concentration phase diagram at band separation
  $\varepsilon_{AB}=8$ and disorders $W_A=W_B=4$. The shaded regions
  denote extended states. The upper region of extended states has been
  constructed from point symmetry with respect to $\varepsilon_{AB}/2$
  and $x_{A}=0.5$. The thick solid lines and the triangle indicate the
  inferred concentration dependencies of 4 different AMA as discussed
  further in section \protect\ref{sec-discussion}.  }
\end{center}
\end{figure}

Obviously, the values for $\Lambda_M$ and the mobility edges should
strongly depend on the concentration.  This is indeed the case as we
show in Fig.~\ref{fig-Ab-Lambda_E}. Already a $10\%$ difference in atomic
composition leads to a pronounced asymmetry of the $\Lambda_M$ curves
(compare, e.g., with Fig.~\ref{fig-Lambda_E}).
The energy-concentration phase diagram for $\varepsilon_{AB}=8$,
and $W_A=W_B=4$ is presented in Fig.~\ref{fig-phase-E_CA}.  For
$x_A<0.32$ ($x_B>0.68$), all electronic states of subband $A$
($B$) are localized.  Figs.~\ref{fig-Ab-Lambda_E} and
\ref{fig-phase-E_CA} correspond to the case of overlapping bands
such that the center of subband $A$ coincides with the edge of
subband $B$ and vice versa.

\subsection{Concentration-$\varepsilon_{AB}$ phase diagram}
\label{sec-conc-EAB}

\begin{figure}
\begin{center}
  \includegraphics[width=0.95\columnwidth]{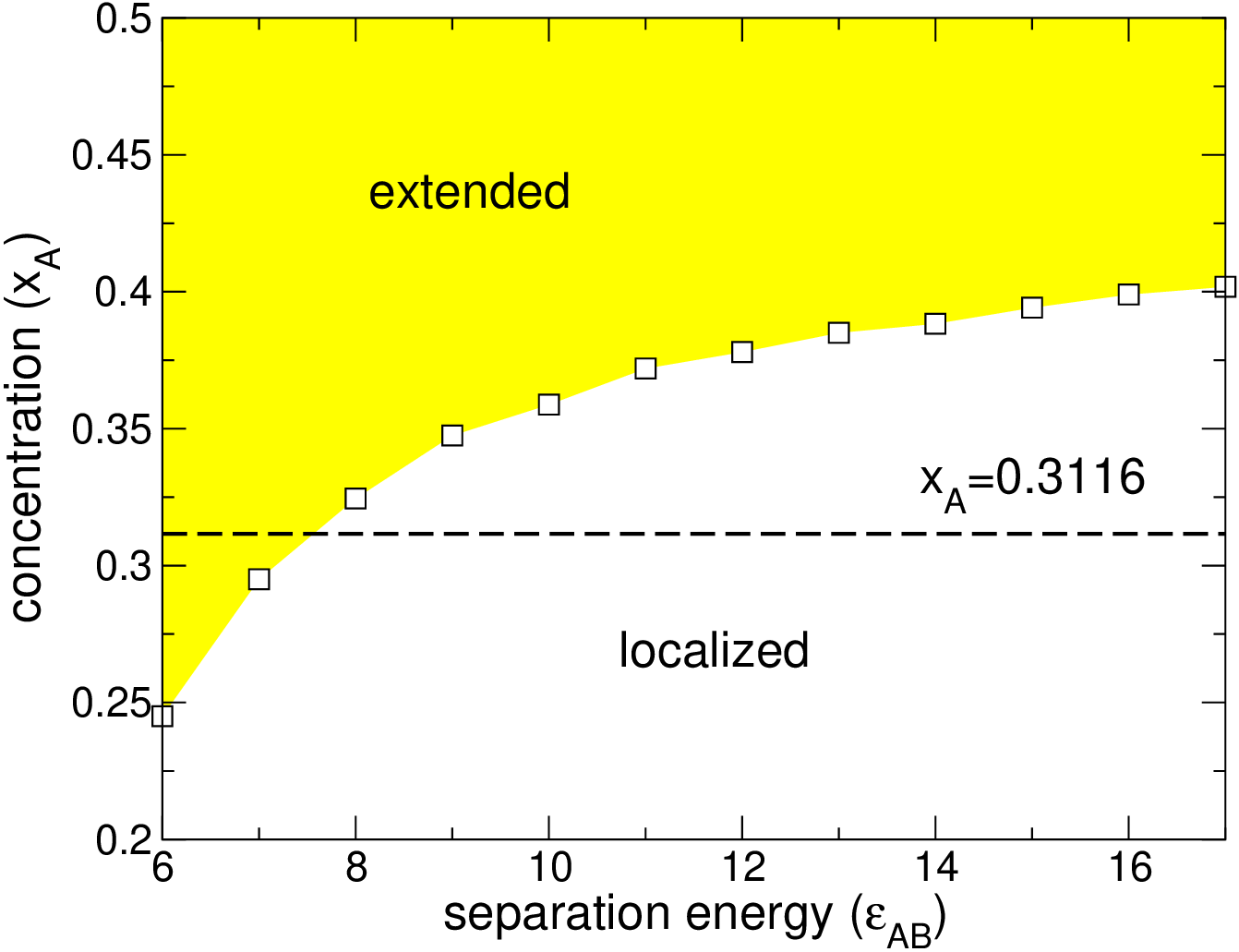}
\caption{\label{fig-phase-CA_EAB}
  The concentration-$\varepsilon_{AB}$ phase diagram at
  $E=\varepsilon_A=0$ and $W_A=W_B=4$. The shaded region denotes
  extended states. The dashed line represents the classical 3D
  site-percolation threshold value $x_{A}=0.3116$.}
\end{center}
\end{figure}

The concentration-$\varepsilon_{AB}$ phase diagram is presented in
Fig.~\ref{fig-phase-CA_EAB} for $E=\varepsilon_A=0$ and
$W_A=W_B=4$.
We see that the critical concentration is larger than the
classical percolation threshold $x_{A}=0.3116$ for large
$\varepsilon_{AB}$. We interpret this as being due to the
localization effect in this quantum situation: a larger cluster of
connected sites is needed to support extended states on sites of
the sublattice $A$ (or $B$). For very large $\varepsilon_{AB}=100$
and $10^3$ (not shown) the critical concentration decreases to
$0.416$ and $0.384$, respectively, but remains larger than in the
classical case. For large subband spacing sites with onsite
potential $\varepsilon_{B}$ can be viewed as potential barriers to
electrons from subband $A$ and vice versa. Thus electrons with
energy $\approx \varepsilon_{A}$ have a low probability to hop
onto $B$ sites and we are effectively studying the case of
percolation.  But even though $\varepsilon_{AB}$ might be large,
there will always remain a finite but small probability for
tunneling thus distinguishing our model from {\em classical}
percolation.  And since we are studying the wave equation
(\ref{eq-Hand}), localization effects not present in classical
percolation are important and lead to insulating behavior even
when the concentration of sites supports a classically percolating
cluster. Therefore, we find that the threshold for transport
(extended states) in the quantum case is higher than for the
classical one. We note that similar effects have been observed
previously in studies of quantum percolation \cite{AviL92}.

On the other hand, for small $\varepsilon_{AB}$, the critical
concentration becomes less than the percolation threshold since
tunneling between $A$ and $B$ sites is more significant. The phase
diagram in Fig.~\ref{fig-phase-CA_EAB} does not exactly show the
mobility edges at which all extended states disappear for a given
$\varepsilon_{AB}$, because it is determined at $E=0$. But as shown in
Figs.~\ref{fig-Lambda_E} and \ref{fig-phase-E_CA}, the $A$ band is not
symmetric with respect to $\varepsilon_A (=0)$. As a consequence, more
extended states occur for energies just below $\varepsilon_A$. This
means that the phase boundary in Fig.~\ref{fig-phase-CA_EAB} will be
shifted slightly downwards, if $E$ is varied, too.

We note that an analogous phase diagram has been obtained
recently~\cite{NazBR02} for the case of a two-dimensional binary
alloy. The shape of the curve separating localized and extended
electronic states is quite similar to our result shown in
Fig.~\ref{fig-phase-CA_EAB}.

\subsection{Disorder-concentration phase diagram}

\begin{figure}
\begin{center}
  \includegraphics[width=0.95\columnwidth]{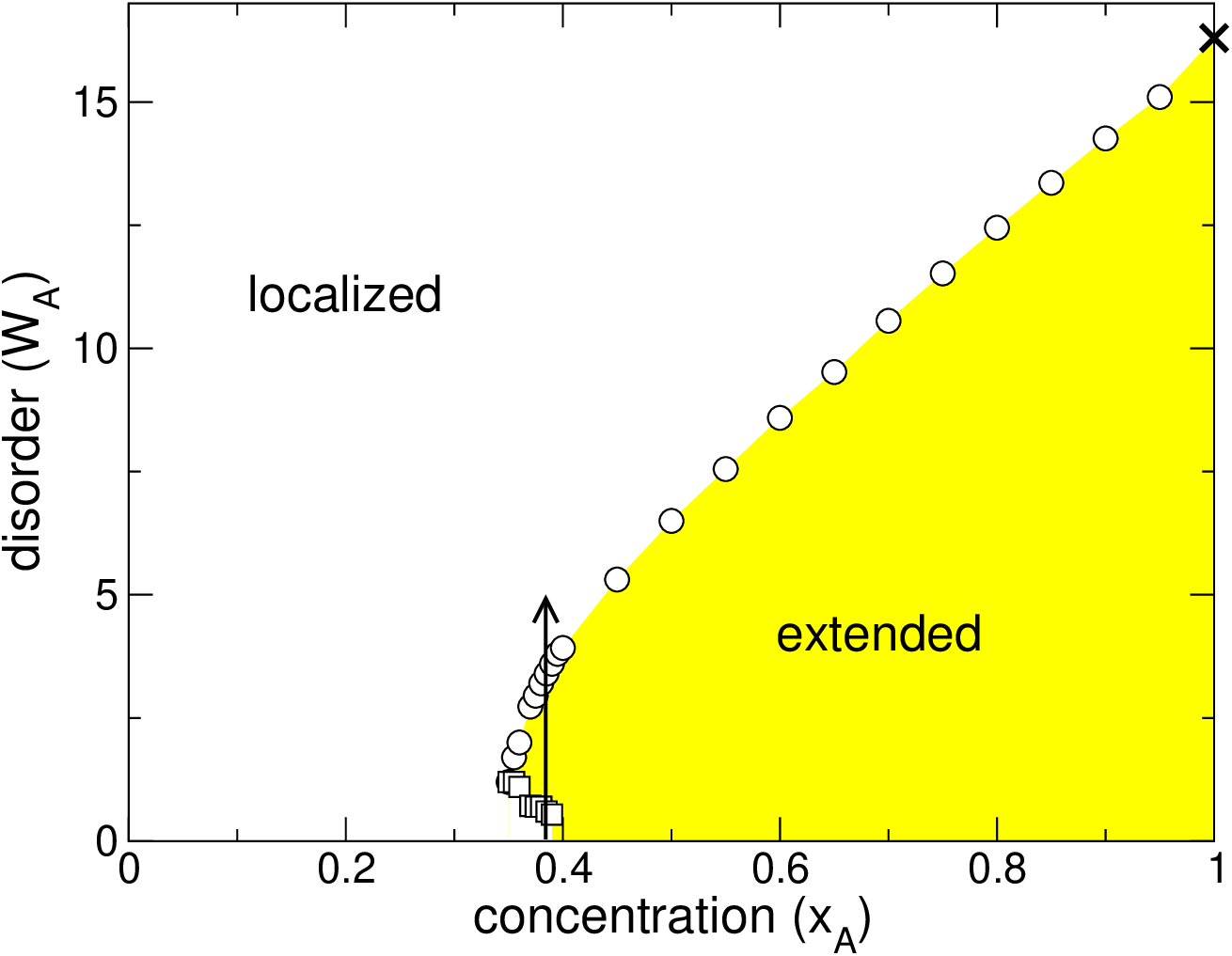}
\caption{\label{fig-phase-WA_CA}
  The disorder-concentration phase diagram for $\varepsilon_{AB}=17$,
  $W_B=4$, and $E=\varepsilon_{A}=0$. The shaded region denotes extended
  states.  The arrow refers to the discussion in the text as well as to
  the data presented in Fig.\ \ref{fig-Lambda_WA}. The $\times$ denotes
  the single-band result.}
\end{center}
\end{figure}

The concentration dependence of the critical disorder --- the
disorder at which all states for a given energy become localized
--- is presented in Fig.~\ref{fig-phase-WA_CA} for energy
$E=\varepsilon_A$.  One can see that the critical disorder
strongly depends on the concentration. For $x_A=1$, we recover the
result $W_{\rm c}=16.5$ \cite{KraM93} of the single-band Anderson
model.
No extended electronic states are observed for $x_{A}<0.35$ in
agreement with the idea that for small concentration the
electronic states of these atoms form the usual localized donor
subband of separated impurities.

\begin{figure}
\begin{center}
  \includegraphics[width=0.95\columnwidth]{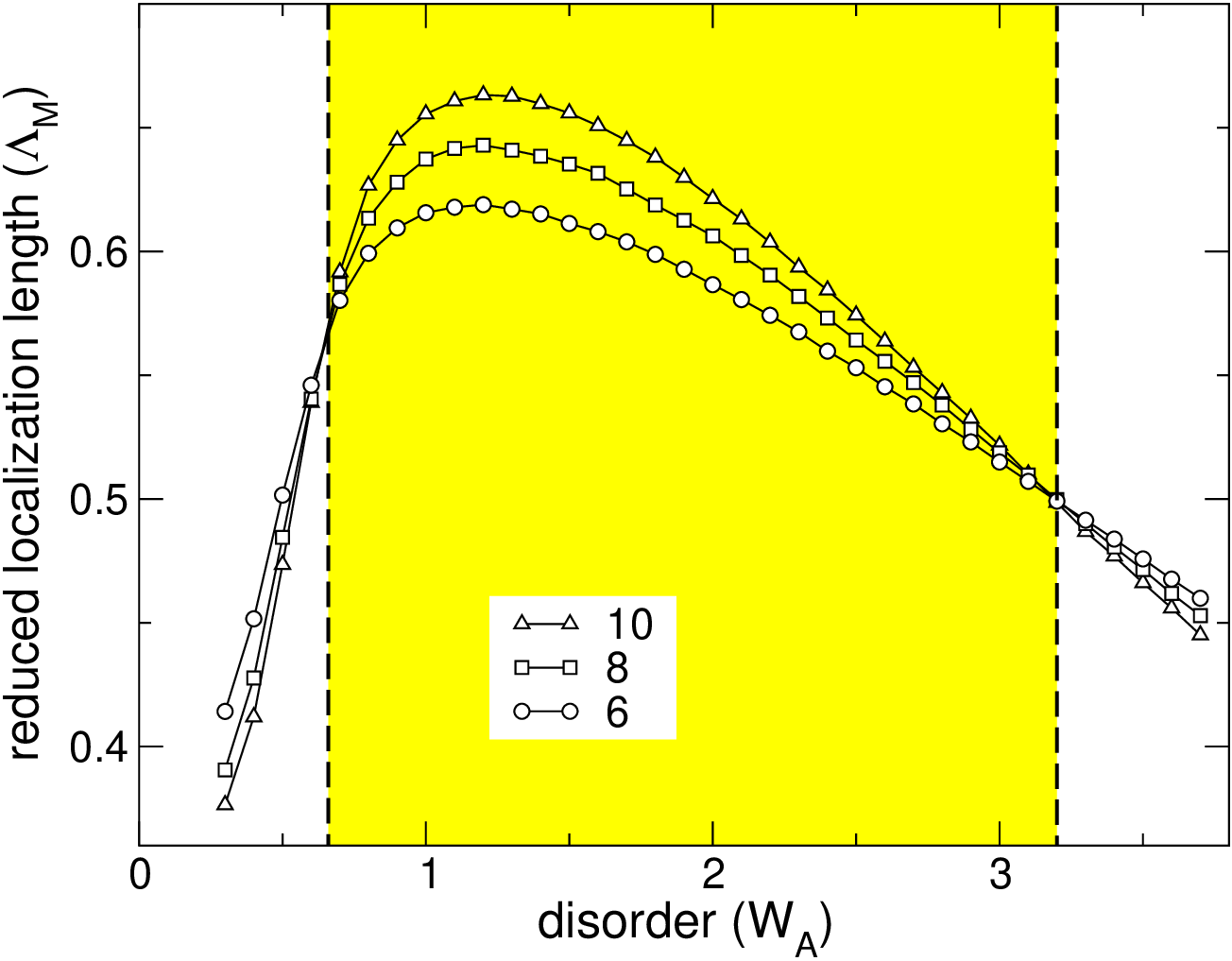}
\caption{\label{fig-Lambda_WA}
  The disorder dependence of $\Lambda_M$ for system sizes $M=6, 8$ and
  $10$ at $\varepsilon_{AB}=17$, $W_B=4$, $E=0$ and $x_{A}=0.38$. The
  shaded region indicates extended states and the two vertical lines
  denote approximate estimates of critical disorders $W_{\rm c}$.}
\end{center}
\end{figure}

For the concentration interval from $0.35$ to $0.40$ we obtain an
additional transition as a function of disorder at small $W_A$
besides the usual Anderson MIT.  For example, as shown in Fig.\
\ref{fig-Lambda_WA} for $x_{A}=0.38$ we observe localized behavior
of $\Lambda_M(M)$ for small disorders. Increasing the disorder
beyond $W_A>0.7$, we see  extended states. At $W_A \sim 3.2$, the
character of the states changes  back to localized.
%
This behavior can be understood as follows: for $x_A\approx 0.38$ as
indicated by the arrow in Fig.\ \ref{fig-phase-WA_CA}, the system at
$W_A=0$ represents a binary alloy with large energy separation
$\varepsilon_{AB}=17$. All states are localized, i.e., non-quantum
percolating, on an already classically percolating set of sites
$\{i(A)\}$ in the $A$ band. Small additional potential disorder will
lead to an increase of certain $\varepsilon_{i(A)}$ as well as a
decrease in certain $\varepsilon_{i(B)}$. A small change of the
scattering conditions in these sites can then quickly allow enhanced
transport across a now quantum-percolating backbone. But upon further
increasing the disorder, the localization in each band will quickly lead
to localization again.

\subsection{Value of the critical exponent}

The critical exponent for the present model can be estimated by
crossing the mobility edges as function of either disorder, energy
or concentration, giving rise to the three exponents $\nu_W$,
$\nu_E$ and $\nu_x$, respectively. Due to universality,
their values should coincide with the critical exponent
$\nu\approx1.6$ of the single-band Anderson model since the
universality class remains unchanged by the introduction of the
binary disorder \cite{HofS93b}.

\begin{figure}
\begin{center}
  \includegraphics[width=0.95\columnwidth]{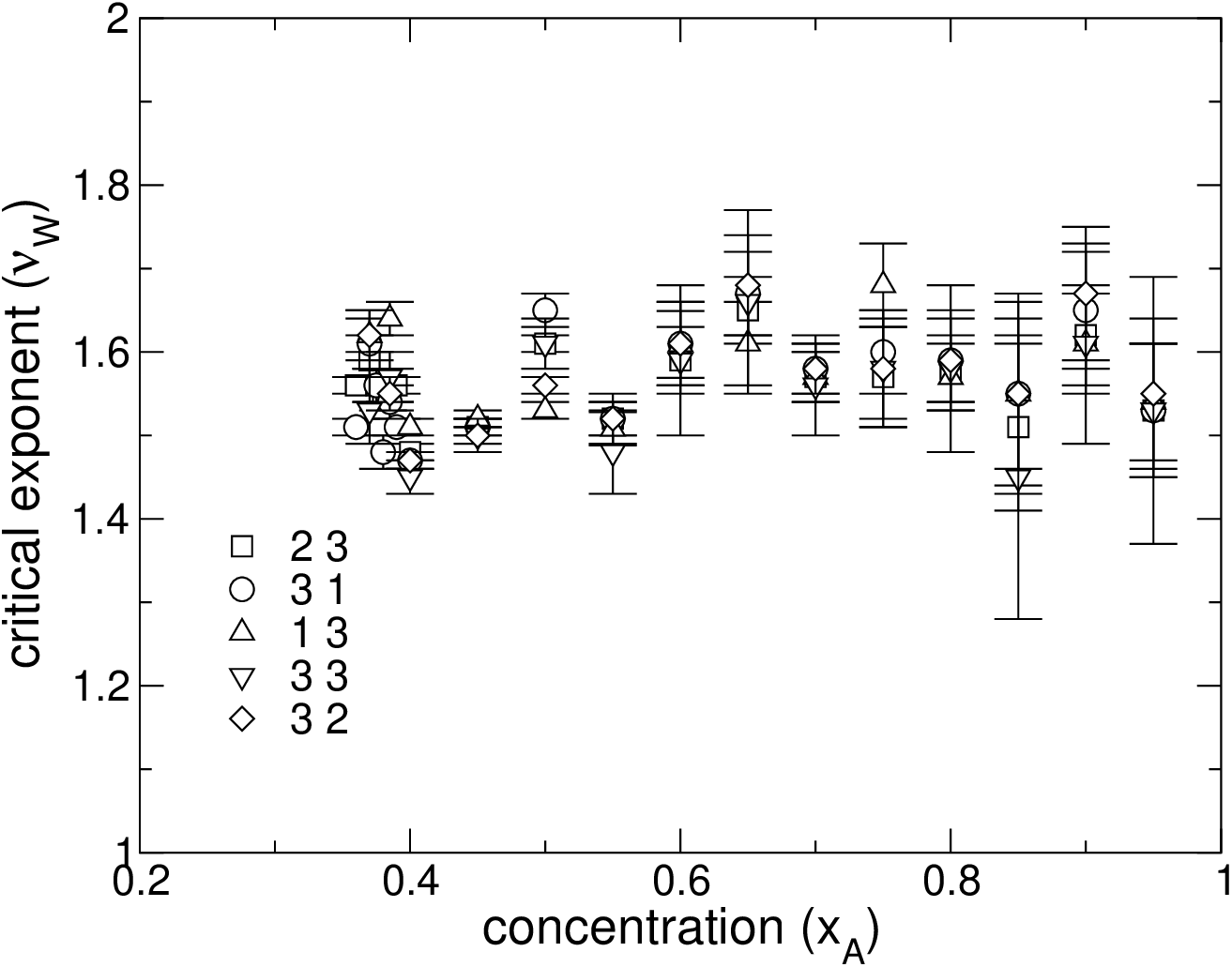}
\caption{\label{fig-nu_CA}
  Concentration dependence of the critical exponent $\nu_W$ obtained
  when varying the disorder across the mobility edge of Fig.\
  \protect\ref{fig-phase-WA_CA}, i.e., for $E=\varepsilon_A=0$,
  $\varepsilon_{AB}=17$ and $W_B=4$. The different symbols represent
  various orders $n_r$ and $m_r$ (indicated in the legend) of the expansion
  (\protect\ref{eq-SlevenRenorm2}) used for the fitting.  The error bars denote one standard
  deviation as obtained from the non-linear fitting procedure.}
\end{center}
\end{figure}

We have estimated the critical exponent $\nu$ for almost all points in
the phase diagrams \ref{fig-phase-E_WA}, \ref{fig-phase-E_EAB},
\ref{fig-phase-E_CA}, \ref{fig-phase-CA_EAB} and \ref{fig-phase-WA_CA}
as described in  section \ref{sec-FSS}.
For example, we compute $\nu_W$ along the mobility edge of the
disorder-concentration phase diagram of Fig.~\ref{fig-phase-WA_CA}. In
Fig.~\ref{fig-nu_CA} we show the resulting estimates when using
different orders $n_r$ and $m_r$ for the fit function (\ref
{eq-SlevenRenorm2}). As usual, the spread in values is somewhat larger
than the least-square-error bars seem to suggest \cite{MilRS00}.
Summarizing the results, we find $\nu_W=1.56\pm0.06$,
$\nu_E=1.64\pm0.05$ and $\nu_x=1.60\pm0.07$, compatible with
$\nu\approx1.6$ \cite{MilRSU00}.

\section{Discussion and comparison with experiments}
\label{sec-discussion}

Let us now compare the above results of the binary Anderson model
with experimental measurements of transport properties of AMA
\cite{NakPSZ00,Cot76}. Such a comparison will of course be purely
qualitative. Nevertheless, it already suffices to understand much
of the transport properties of such AMA. E.g., as mentioned in
section \ref{sec-intro}, it is the position of the mobility edge
relative to the value of the Fermi energy that determines the
properties of transition-metal-based AMA.  In this spirit, the
Mooji correlation \cite{Moo73} is due to $E_{\rm F}$ being
situated in a region of localized states, whereas the weak changes
in resistivity at melting for Fe and Co as well as the positive
temperature coefficient of resistivity for Fe-based AMA result
from $E_{\rm F}$ being located in the center of the 3d-electronic
band and hence, in the region of extended states.

\subsection{Transport properties of Ni$_{1-x}$P$_x$}

The ionization energy of P is higher than that of Ni, so our $A$
band  corresponds to P atoms and the $B$ band to Ni atoms with
$\varepsilon_B>\varepsilon_A$ as, e.g., shown in Fig.\
\ref{fig-phase-E_CA}. According to the theoretical and
experimental results of Ref.~\cite{BabPSZ95}, $E_{\rm F}$ of the
amorphous alloy Ni$_{1-x}$P$_x$ is located near the upper edge of
the Ni-3d band (the $B$ band in our study). When the P
concentration $x_{\rm P}$ increases from $0.15$ to $0.27$ the
temperature coefficient of resistivity (TCR)
 decreases from positive to negative values
\cite{Cot76}. The MIT is observed near  $x_{\rm P}=0.25$.
In Fig.\ \ref{fig-phase-E_CA}, we indicated the corresponding
region in the phase diagram. The two-band, binary random potential
Anderson model can indeed qualitatively capture the observed
experimental results. Even a quantitative comparison might be
possible if a reliable way of obtaining values for
$\varepsilon_{A}$, $\varepsilon_B$, and $W_A$ and $W_B$ becomes
known. Note, however, that in the experiments, it is the hopping
between $A$ and $B$ atoms at their respective random positions
that leads to the broadening of the $A$ and $B$ bands and we have
modeled this only qualitatively in the present work by the onsite
disorders $W_A$ and $W_B$.

\subsection{Transport properties of Ti$_{1-x-y}$Cu$_x$A$_y$ and Ti$_{1-x-y}$Ni$_x$A$_y$}

A previous experimental study of the electronic structure of
Ti$_{1-x-y}$Cu$_x$A$_y$ and Ti$_{1-x-y}$Ni$_x$A$_y$ amorphous alloys
\cite{NakPSZ00} (the symbol A$_y$ represents additional admixtures) has
shown that the valence band has two main peaks. The lower peak is formed
predominantly by the 3d states of Cu or Ni whereas the upper peak is due
to the 3d states of Ti. The Fermi level is located in the central region
between the peaks and is hardly shifted upon changing the
concentrations.

Within the present two-band model, this experimental situation may be
modeled as shown in Fig.\ \ref{fig-phase-E_CA}. The line for the
Ti$_{1-x-y}$Ni$_x$A$_y$ alloys is situated somewhat below the one for
Ti$_{1-x-y}$Cu$_x$A$_y$ because Ni has one electron less than Cu. We believe
that it is this difference that leads to the smaller value of the
negative TCR for Ti$_{1-x-y}$Ni$_x$A$_y$ when compared to Ti$_{1-x-y}$Cu$_x$A$_y$. In
both cases, $E_{\rm F}$ lies in a region of localized states in
agreement with the negative TCR. In Table \ref{tab-TCRexp}, we show the
values obtained for TCR.
\begin{table}
\caption{\label{tab-TCRexp}
  Temperature coefficient of resistivity $\alpha_{300}$ at $300$ K for
  Ti$_{1-x-y}$Cu$_{x}$A$_y$ and Ti$_{1-x-y}$Ni$_{x}$A$_y$ amorphous alloys \cite{NakPSZ00}. }
\begin{center}
\begin{tabular}{l|c|l|c}
\hline
\multicolumn{1}{c|}{AMA} &$\alpha_{300}$  & \multicolumn{1}{c|}{AMA} & $\alpha_{300}$ \\
    &$10^{-4}/$K &     & $10^{-4}/$K \\
\hline
Ti$_{62}$Cu$_{33}$P$_{5}$ & -0.70 & Ti$_{70}$Ni$_{25}$Si$_{5}$ & -2.21 \\
Ti$_{47}$Cu$_{45}$Ni$_{5}$Si$_{3}$ & -1.11 & Ti$_{60}$Ni$_{36}$P$_{2}$Si$_{2}$ & -2.72 \\
Ti$_{48}$Cu$_{45}$Ni$_{5}$P$_{2}$ & -1.05 & Ti$_{50}$Ni$_{45}$P$_{5}$ & -2.37 \\
Ti$_{46}$Cu$_{45}$Ni$_{5}$Si$_{2}$P$_{2}$ & -3.03 & Ti$_{45}$Ni$_{50}$P$_{5}$ & -3.28 \\
\hline
\end{tabular}
\end{center}
\end{table}

Let us now analyze the concentration dependence of the TCR. When
the Ti concentration decreases, the TCR also decreases as shown in
Table \ref{tab-TCRexp}. This behavior can be explained from the
concentration-energy phase diagram of Fig.~\ref{fig-phase-E_CA} by
the increase of the mobility edge $E_{\rm c}$. This in turn leads
to an increasing distance between $E_{\rm F}$ and the region of
unoccupied extended electronic states. The states close to $E_{\rm
F}$ become more localized, leading to a more negative TCR.

Therefore, at least at a qualitative level, the present two-band
Anderson model explains not only the negative TCR of
Ti$_{1-x-y}$Cu$_x$A$_y$ and Ti$_{1-x-y}$Ni$_x$A$_y$ glasses but also the
changes in TCR due to changes in composition (Cu or Ni) and
concentration.  Furthermore, metallic glasses with a ``metal''-like
conductivity can be also treated within this model.  For example, the
Fermi level of Fe$_{86}$B$_{14}$ alloy is situated in the region of
extended states as indicated in Fig.~\ref{fig-phase-E_CA} in agreement
with its positive TCR \cite{NakPSZ00}.

\subsection{Overlapping the two bands}

Thus far, motivated by the experimental results, we concentrated on the
case of separated bands, i.e., large $\varepsilon_{AB}$. To make the two
regions of extended states in Fig.~\ref{fig-phase-E_CA} overlap, we can
in principle (i) keep $\varepsilon_{AB}$ fixed and change the disorder,
or (ii) keep fixed disorder and decrease $\varepsilon_{AB}$ as in
Fig.~\ref{fig-phase-E_EAB}.  From Fig.\ \ref{fig-phase-E_WA}, we know
that the position of mobility edges does not change a lot when
decreasing $W_A$. On the other hand, increasing $W_A$ will lead to a
narrower region of extended states in Fig.\ \ref{fig-phase-E_CA} and no
overlap.  Therefore, we choose to reduce $\varepsilon_{AB}$ in order to
study what happens when the two bands overlap in the
energy-concentration diagram.

\begin{figure}
\begin{center}
  \includegraphics[width=0.95\columnwidth]{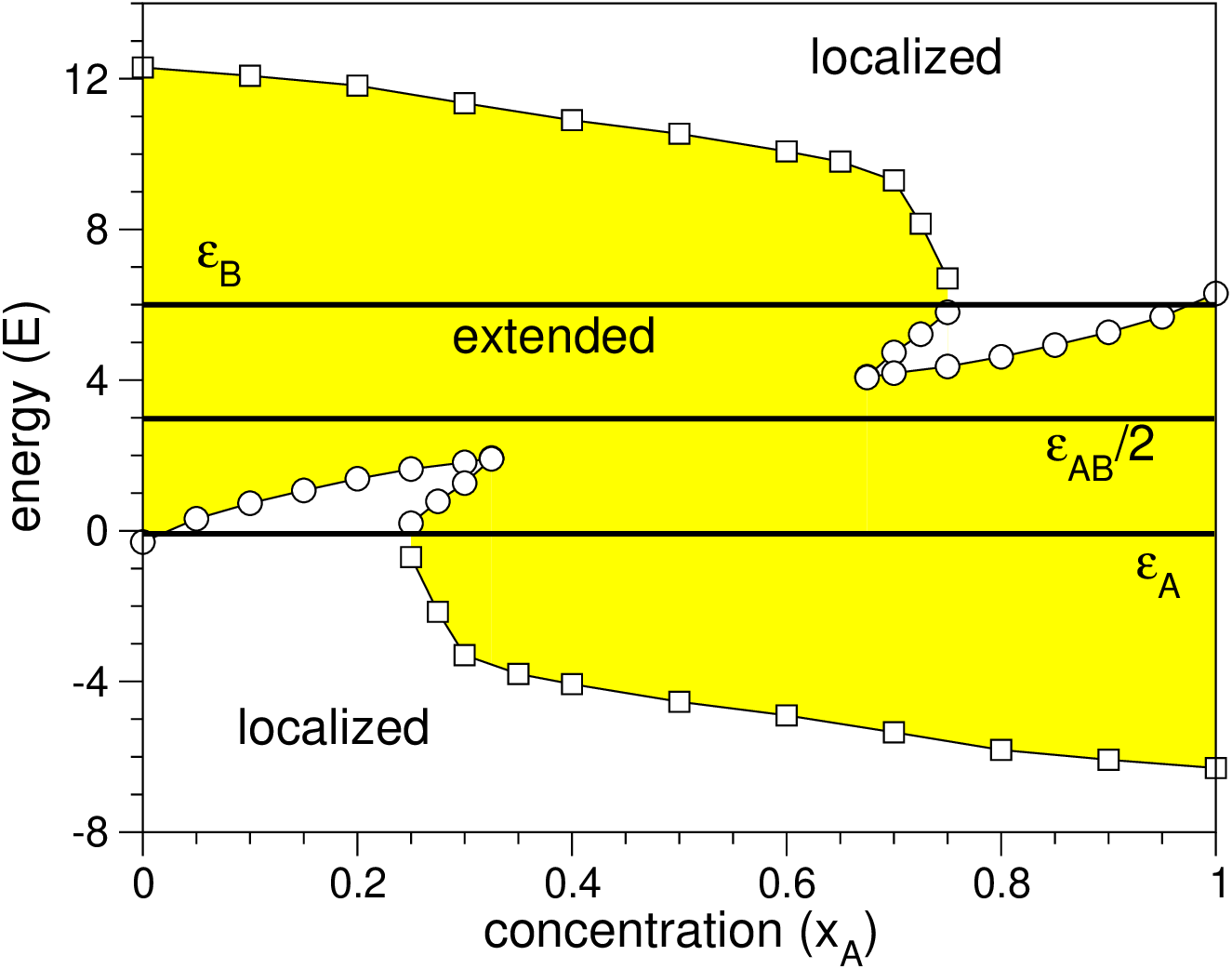}
\caption{\label{fig-phase-E_CA-overlap}
  The energy-concentration phase diagram at band separation
  $\varepsilon_{AB}=6$ and disorders $W_A=W_B=4$. The shaded region denotes
  extended states. The horizontal lines indicate $\varepsilon_B$,
  $\varepsilon_{AB}/2$ and $\varepsilon_{B}$ from top to bottom, respectively.
  The upper mobility edge has been constructed from point symmetry
  with respect to $\varepsilon_{AB}/2$ and $x_{A}=0.5$.}
\end{center}
\end{figure}

In Fig.~\ref{fig-phase-E_CA-overlap}, we show such an overlap for
$\varepsilon_{AB}=6$ and $W_A=W_B=4$.
This situation corresponds to  large overlap of the two subbands.
We see that the shape of the regions of extended states remains
nearly unchanged. This leads to non-monotonicity in the behavior
of the upper and lower mobility edges. We remark that in order to
resolve this behavior, the accuracy of the TMM had to be increased
significantly up to very small errors of $0.05\%$ for the
localization lengths. We emphasize that a similar accuracy was
needed to resolve the non-monotonicity in the
disorder-concentration diagram in Fig.\ \ref{fig-phase-WA_CA}.

When further decreasing $\varepsilon_{AB}$ to $5$, the behavior of
the mobility edges becomes regular again and their values
eventually reduce to the values of the well-known phase diagram of
the single-band Anderson model \cite{BulKM85}.

\section{Conclusions}

We have shown that the three-dimensional Anderson model of
localization with binary random potential disorder allows to
explain not only various peculiarities of the
transition-metal-based amorphous alloys but also how these change
as the chemical composition changes.
Even for a comparatively large overlap of the two subbands, a
region of localized electronic states exists in the central part
between the two subbands. This confirms the assumption made in
Ref.\ \cite{NakPSZ00} for the experimental results in AMA. Even a
quantitative analysis, using suitably extracted parameters from
density-of-states measurements appears possible. Of course in this
case ternary and quaternary AMA will have to be distinguished by
additional onsite potentials similar to $\varepsilon_{A}$ and
$\varepsilon_{B}$ for the binary model.

In addition to the usefulness of the two-band Anderson model for a
qualitative comparison with experiments, there is also the interesting
fact that the model allows to study the concentration dependence and the
associated MIT. This concentration dependence is clearly a prominent
feature present in most disordered materials. We find that, as expected
from universality, the universality class of this additional transition
remains unchanged with respect to a single-band Anderson model.

\begin{acknowledgments}

We thank A.\ Eilmes, S.\ Klassert and P.\ Stollmann for useful
comments. This work was supported by grants from the DAAD via
Ref.\ 322, A/01/20012 and the DFG via SFB393 .
\end{acknowledgments}

\vspace*{3ex}

\end{document}